# Superconductivity in Uncollapsed Tetragonal LaFe$_2$As$_2$

Akira Iyo[1], Shigeyuki Ishida[1], Hiroshi Fujihisa[1], Yoshito Gotoh[1], Izumi Hase[1], Yoshiyuki Yoshida[1], Hiroshi Eisaki[1], Kenji Kawashima[1,2]

[1]National Institute of Advanced Industrial Science and Technology (AIST), 1-1-1 Umezono, Tsukuba, Ibaraki 305-8568, Japan

[2]IMRA Material R&D Co., Ltd., 2-1 Asahi-machi, Kariya, Aichi 448-0032, Japan

**ABSTRACT:** We report synthesis, crystal structure and superconductivity in ThCr$_2$Si$_2$-type LaFe$_2$As$_2$ (La122). La122 was synthesized at 960°C for 1.5 h under a pressure of 3.4 GPa. An as-synthesized La122 (non-superconductor) had a collapsed tetragonal structure with a short $c$-axis length of 11.0144(4) Å as observed in CaFe$_2$As$_2$ under pressure. The collapsed tetragonal transformed into an uncollapsed tetragonal by annealing the as-synthesized La122 at 500°C. The $c$-axis length remarkably extended to 11.7317(4) Å and superconductivity emerged at 12.1 K in the uncollapsed tetragonal La122. A cylindrical hole-like Fermi-surface around the $\Gamma$ point that plays an important role for an $s\pm$ wave paring in iron-based superconductors was missing in the uncollapsed tetragonal La122 due to heavily electron-doping. Superconductivity in La122 may be closely related to that induced in CaFe$_2$As$_2$ under pressure.

High-$T_c$ superconductivity in iron arsenides is typically induced by doping holes or electrons into non-superconducting parent compounds such as CuZrSiAs-type $Ln$FeAsO ($Ln$1111) and ThCr$_2$Si$_2$-type $Ae$Fe$_2$As$_2$ ($Ae$122), where $Ln$ and $Ae$ are lanthanoid and alkaline earth metal elements, respectively. $Ln$1111 is suitable for electron-doping and a superconducting transition temperature ($T_c$) above 50 K is realized through a partial substitution of F$^-$ or H$^-$ for O$^{2-}$,[1–6] or by introducing an oxygen deficiency.[7,8] $Ae$122 favors hole-doping, and superconductivity up to 38 K appears when approximately half of $Ae^{2+}$ is substituted for with $A^+$ ($A$ is an alkali metal element).[9]

On the other hand, hole-doping on $Ln$1111 and indirect electron-doping on $Ae$122 are difficult to realize, especially when using conventional solid-state reaction methods. Up to 25% of $Ln^{3+}$ in $Ln$1111 is replaced by Sr$^{2+}$ or Ca$^{2+}$ (hole-doping) and superconductivity occurs at 13.5–25 K.[10,11] For $Ae$122, $Ae^{2+}$ is replaced by $Ln^{3+}$ up to 25% (electron-doping), and the onset $T_c$ (~49 K) is higher than the hole-doped $Ae$122.[12–16]

The electron-doping range in $Ae$122 is expanded using a high-pressure (HP) or thin-film synthesis method.[17–18] (Sr$_{1-z}$La$_z$)Fe$_2$As$_2$ is synthesized under HP for $0 \leq z \leq 0.5$, which corresponds to a Fe formal valence ($v_{Fe}$) range of $+1.75 \leq v_{Fe} \leq +2.0$; superconductivity at 22 K is found for $x = 0.4$. However, further electron-doping ($v_{Fe} < +1.75$) was not achieved for $Ae$122-type iron arsenides. Note that the situation concerning the doping is different in iron selenides, namely, 122-type iron selenides $A_x$Fe$_{2-y}$Se$_2$ ($A$ = K, Rb, Ce, Tl) are easily electron-doped.[19–21]

For $Ln$1111-type compounds, electron-doped samples with a wider $v_{Fe}$ range of $+1.25 - +1.5 \leq v_{Fe} \leq +2.0$ were synthesized for $Ln$FeAs(O$_{1-y}$H$_y$) ($0 \leq y \leq 0.5$), ThFeAs(N$_{1-y}$O$_y$) ($0 \leq y \leq 0.6$) and LaFeAs(O$_{1-y}$F$_y$) ($0 \leq y \leq 0.75$) under HP. As a result, characteristic properties such as two superconducting domes in a wide-range electronic phase diagram were found.[5,22,23]

From the doping polarity viewpoint, (La$_{0.5-x}$Na$_{0.5+x}$)Fe$_2$As$_2$ ((La,Na)122) is an attractive system because either electrons ($-0.5 \leq x < 0$) or holes ($0 < x \leq 0.5$) can be doped by changing one parameter (composition $x$).[24] In our previous study, we succeeded in synthesizing hole-doped samples ($0 \leq x \leq 0.35$) and found superconductivity at 28 K for $x \sim 0.3$.[25] However, we failed to synthesize electron-doped samples using a conventional synthesis method under normal pressure.

In this paper, we report the successful synthesis of LaFe$_2$As$_2$, which is an end member ($x$ = -0.5) of (La,Na)122, using the HP technique. The as-synthesized sample has a collapsed tetragonal structure as observed in CaFe$_2$As$_2$ (Ca122) under pressure.[26] We found that the structure phase transition from collapsed to uncollapsed tetragonal occurred by annealing samples, which resulted in superconductivity emergence at 12.1 K.

Polycrystalline samples were synthesized using a cubic-anvil-type HP apparatus. Compounds of LaAs and Fe$_2$As were used as starting materials. For example, a sample with a nominal composition of La$_{1.1}$Fe$_2$As$_{2.1}$ was ground using a mortar in an N$_2$-filled glove box. There was a 5–10 at% excess of LaAs, which eventually decreased the composition ratio of Fe but was found to improve sample quality. The ground powder was pressed into a pellet, which was then placed in a BN crucible and assembled into a HP cell.[27] The sample was heated at 960°C for 1.5 h under a pressure of 3.4 GPa, followed by rapid cooling (below 100°C in 1 min). Reaction temperature had to be tuned carefully,[25,28] or else the yield of La122 in a sample decreased substantially. Subsequently, the sample was annealed at 500°C for 10 h in an evacuated quartz tube. Both the as-synthesized and annealed samples were stable in air at least for several months.

Powder X-ray diffraction (XRD) patterns were measured at room temperature using a diffractometer (Rigaku, Ultima IV) with Cu$K_a$ radiation (wavelength of 1.5405 Å). The crystal structures of the samples were refined via a Rietveld analysis using BIOVIA's Materials Studio Reflex software (version 2018 R2).[29] Magnetization ($M$) measurements were performed under a magnetic field ($H$) of 10 Oe using a magnetic-property measurement system (Quantum Design, MPMS-XL7). The electrical resistivity was measured using a four-probe method (Quantum Design, PPMS). Sample compositions were analyzed by an energy dispersive X-ray spectrometry (Oxford, SwiftED3000) equipped in an electron microscope (Hitachi High-Technologies, TM3000). To avoid possible degradation layers on the sample surface, the inner part of the sample was used for the measurements.

*Ab-initio* electronic structure calculations were performed using the full-potential linearized augmented plane wave method in the WIEN2K computer code package.[30] We used the parameter $RK_{max}$ = 7.0, and the muffin-tin radii set as $r$(La) = 2.5 a.u., $r$(Fe) = 2.3 a.u., and $r$(As) = 2.19 a.u. We used the space group, experimental lattice constants, and the atomic positions of the annealed sample.

Figure 1 shows powder XRD patterns of the as-synthesized (CT-La122) and annealed (UT-La122) samples. Most diffraction peaks can be assigned to La122 (a body-centered tetragonal lattice with an $I4/mmm$ space group), as indicated by the closed circles and diffraction indices ($hkl$). An impurity phase of LaFeAsO (La1111) also formed in the samples due to oxidization/hydroxylation of the starting materials.

Structural parameters of CT-La122 and UT-La122 obtained through Rietveld structure refinements are summarized in Table 1. Figure 2 shows the Rietveld fitting and the refined crystal structure with structural parameters for UT-La122.

Lattice parameters of CT-La122 were found to be $c$ = 11.0144(4) Å and $a$ = 4.0035(1) Å. From apparent peak shifts (see e.g., 200 and 116 peaks) in Fig. 1, it is evident that lattice parameters of UT-La122 drastically changed to $c$ = 11.7317(4) Å and $a$ = 3.9376(1) Å, while keeping the same space group. At the same time, the diffraction peak intensity of the LaAs impurity phase increased in UT-La122, as indicated by the open triangles in Fig. 1.

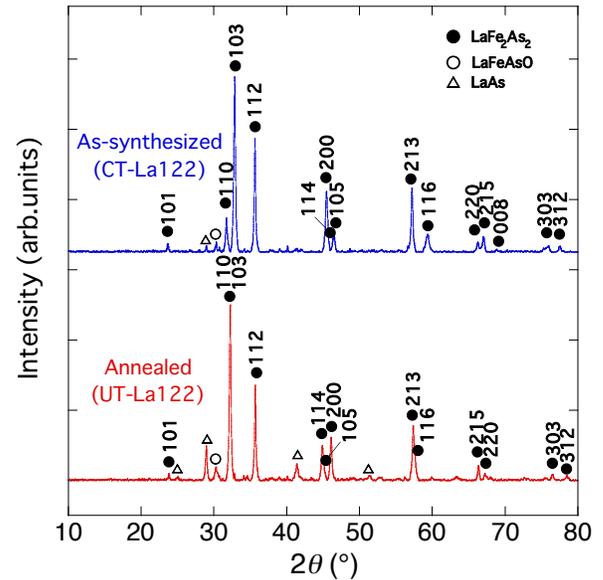

Figure 1. Powder XRD patterns of CT-La122 and UT-La122. Backgrounds were subtracted from the data.

**Table 1. Results of Rietveld structure refinements for CT-La122 and UT-La122 at room temperature**

| Samples | CT-La122 | UT-La122 |
|---|---|---|
| $c$ (Å) | 11.0144(4) | 11.7317(4) |
| $a$ (Å) | 4.0035(1) | 3.9376(1) |
| $V$ (Å$^3$) | 176.5(1) | 181.9(1) |
| $z_{As}$ | 0.3589(1) | 0.3657(1) |
| $l_{Fe-As}$ (Å) | 2.334(2) | 2.391(2) |
| $l_{As-As}$ (Å) | 3.108(2) | 3.151(2) |
| $\alpha_{As-Fe-As}$ (°) | 118.1(2) | 110.8(2) |
| $h_{As}$ (Å) | 1.200(2) | 1.357(2) |
| $R_{wp}$ (%) | 10.88 | 9.42 |
| $R_e$ (%) | 9.38 | 6.75 |
| $S$ | 1.16 | 1.4 |

Space group: $I4/mmm$. The occupancy was set to 1 at all atomic sites. Atomic coordination of La, Fe, and As are (0, 0, 0), (1/2, 0, 1/4), and (0, 0, $z_{As}$), respectively. Global isotropic temperature factors were employed. The mean-square displacements $U_{iso}$ (Å$^2$) of CT-La122 and UT-La122 were 0.048(1) and 0.037(1), respectively.

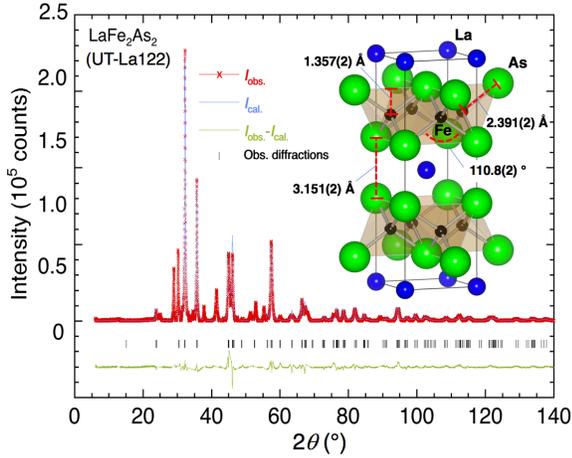

Figure 2. Rietveld fitting and refined crystal structure for UT-La122, illustrated using VESTA software.[31] The sample batches used for the Rietveld analysis were different from those shown in Fig. 1. The 2θ regions of impurity peaks were excluded in the analyses. $I_{obs.}$ and $I_{cal.}$ represent the observed and calculated diffraction intensities, respectively.

Compositions were analyzed using point measurements on the polished surfaces of CT-La122 and UT-La122 pellets. In addition to the La122 phase, spots with composition ratios corresponding to La1111 and LaAs phases were detected. The composition ratios of La122 were determined by averaging measured values of approximately 50 spots. The obtained composition ratios of La: Fe: As were 1.02 (3): 1.80 (4): 2.00 (4) for CT-La122 and 1.01 (3): 1.94 (4): 2.00 (4) for UT-La122. CT-La122 had a Fe deficiency, while UT-La122 had a nearly stoichiometric composition ratio (La: Fe: As = 1: 2: 2).

Figure 3 shows the temperature ($T$) dependence of $4\pi M/H$ for CT-La122 and UT-La122. While CT-La122 exhibited only traces of superconductivity, a large superconducting (diamagnetic) transition surprisingly appeared for UT-La122 with an onset $T_c$ of 12.1 K (see inset in Fig. 3). A shielding volume fraction, approximately 100% at 2 K, was large enough to be regarded as a bulk superconductor.

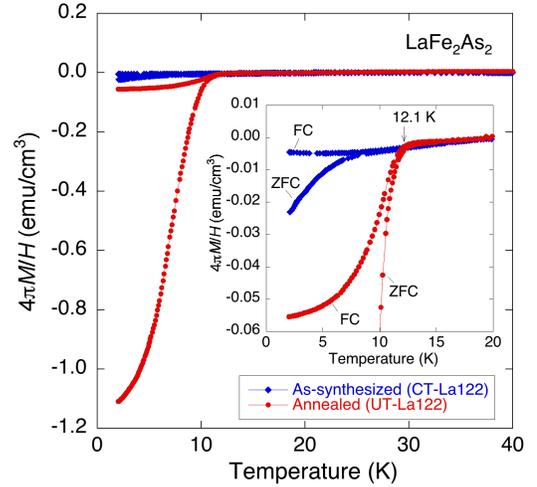

Figure 3. $T$ dependence of $4\pi M/H$ for CT-La122 and UT-La122. Measurements were performed with zero-field-cooling (ZFC) and field-cooling modes. The inset shows the enlargement of transitions.

Figure 4 shows the $T$ dependence of resistivity for CT-La122 and UT-La122. The anomaly that generally appears in iron-based compounds, along with structure phase transitions such as tetragonal-orthorhombic[9] and tetragonal-collapsed tetragonal,[32] was not detected for either sample.

CT-La122 exhibited weak $T$-dependent resistivity, with a small residual resistivity ratio ($RRR$) of 2.3, where $RRR$ is defined by the resistivity ratio at 300 K and 20 K. A trace of superconductivity observed in resistivity as well as magnetization is likely due to a small amount of superconducting component such as LaFeAsO$_{1-y}$/LaFeAs(O,H) formed under HP[4–8] and/or UT-La122 partially formed in the as-synthesized sample. UT-La122 showed a more metallic $T$- dependent resistivity than CT-La122, with a large *RRR* of 15.5. It had a sharp superconducting transition with an onset $T_c$ of 13.0 K (see inset in Fig. 4), which is close to the value measured by magnetization.

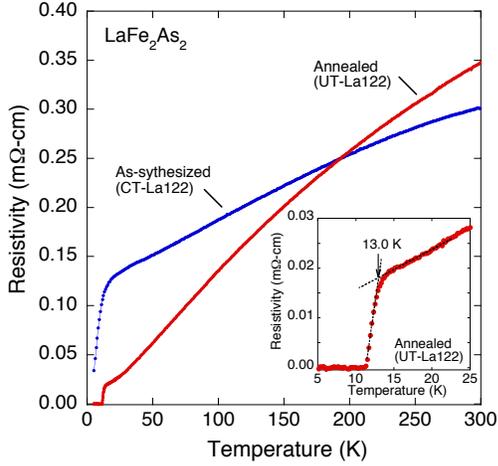

Figure 4. *T* dependence of resistivity for CT-La122 and UT-La122. The inset shows an enlargement near the superconducting transition of UT-La122.

The *T* dependence of the resistivity in a magnetic field up to 9 T is shown in Fig. 5 for UT-La122. $T_c$ decreased as the applied magnetic field increased. The *T* dependence of the upper critical field $H_{c2}$ defined by midpoint $T_c$ is shown in the inset. On the basis of the Werthamer–Helfand–Hohenberg (WHH) equation, $\mu_0 H_{c2}(0) = -0.693 T_c (dH_{c2}/dT)$ for a type-II superconductor in the dirty limit,[33] $\mu_0 H_{c2}(0)$ is estimated at 15.0 T, where $T_c$ (midpoint) = 12.2 K and $dH_{c2}/dT = -1.77$ T/K are used.

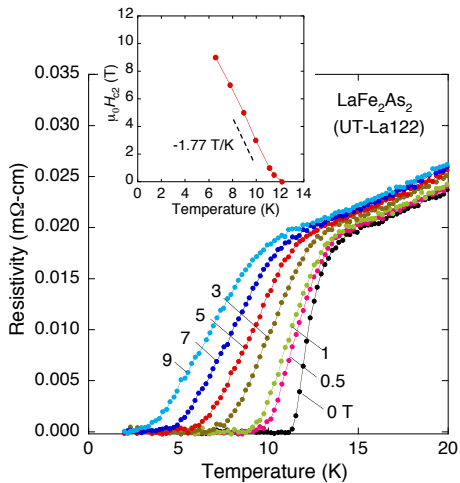

Figure 5. *T* dependence of the resistivity as a parameter of a magnetic field up to 9 T for UT-La122. The upper critical magnetic field $H_{c2}$ defined by the midpoint $T_c$ are plotted against *T* in the inset.

Fermi surfaces obtained through *ab-initio* electronic structure calculations are depicted in Figure 6 for UT-La122. UT-La122 had two electron-like Fermi surfaces at the zone corner, one is cylindrical and the other is jungle-gym-like (Fig. 6d and e, respectively). UT-La122 also had three hole-like Fermi surfaces at the Z point, two of them are small pocket-like (Fig. 6a and b) and the other is disk-shaped (Fig. 6c). It should be noted that the cylindrical hole-like surface around the Γ point, which generally appears in related iron-based compounds, disappeared due to heavy electron-doping. It would be meaningful to compare the Fermi surface topology among related compounds. The Fermi surface topology of UT-La122 is similar to that of the analogous compound LaFe$_2$P$_2$. The cylindrical hole-like Fermi surface is also absent in the collapsed tetragonal Ca122.[34–37]

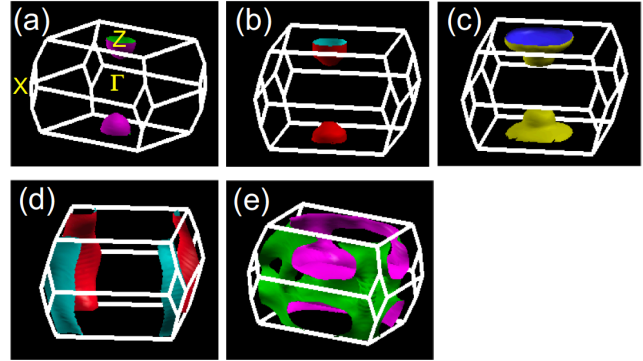

Figure 6. Fermi surfaces in the uncollapsed tetragonal LaFe$_2$As$_2$ (UT-La122).

In Fig. 7, the lattice parameters of CT-La122 and UT-La122 ($x = -0.5$) are plotted together with the lattice parameters of hole-doped (La,Na)122 ($0 \leq x \leq 0.35$). The *c*-axis (*a*-axis) length of CT-La122 is much shorter (longer) than the value expected from extrapolation (Vegard's Law), as shown by the dashed lines in Fig. 7. The arrows in Fig. 7 show that the lattice parameters of UT-La122 steadily approached the expected values. The *c*-axis (*a*-axis) length extended (shrunk) by 0.708 (0.067) Å due to annealing.

The large change in the lattice parameters can be explained by the structure transition between a collapsed and uncollapsed tetragonal reported in the same structure-type compound of Ca122.[26] The *c*-axis ($a$-axis) length shrinks (extends) by 0.755 (0.0709) Å due to the structure collapse at 150–200 K under a pressure of 6.3 GPa. The magnitudes of the lattice parameter changes in La122 are close to those in Ca122. In addition, CT-La122 and the collapsed Ca122 have common structure features in that $\alpha_{As-Fe-As}$ ($h_{As}$) is unusually large (short) ($\alpha_{As-Fe-As}$ = 116.4° and $h_{As}$ = 1.23 Å for Ca122) compared to the other related iron-based compounds.[38,39]

In the collapsed structure, As forms dimers and the valence of As changes from -3 to -2 through dimerization. Assuming that all the As form dimers, $v_{Fe}$ in CT-La122 is +0.5, which is unusually small. According to the structure refinements, $l_{As-As}$ in CT-La122 was 3.108(2), which is longer than that (2.844Å) in the collapsed Ca122.[26] Therefore, we infer that As partially forms dimers in CT-La122 and $v_{Fe}$ is maintained at greater than +1. In contrast, $l_{As-As}$ in UT-La122 (3.151(2) Å) is comparable to that in uncollapsed Ca122 (3.13 Å)[40], which indicates that As is completely dedimerized in UT-La122.

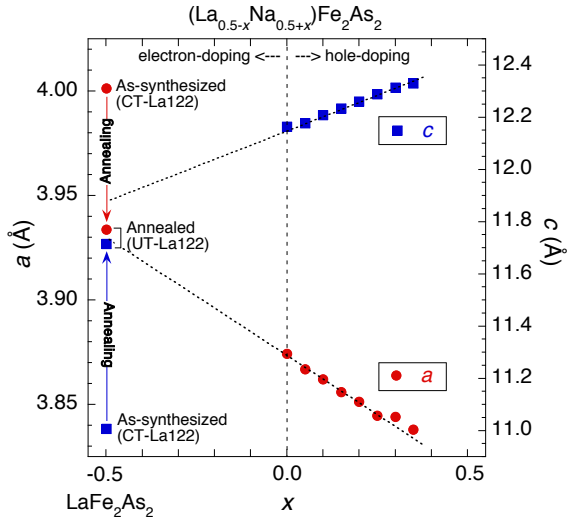

Figure 7. Lattice parameters of CT-La122 and UT-La122 plotted on the composition dependence of lattice parameters for (La$_{0.5-x}$Na$_{0.5+x}$)Fe$_2$As$_2$. Dashed lines indicate the extrapolations of lattice parameters from the hole-doped (La,Na)122 (0 ≤ $x$ ≤ 0.35), based on Vegard's Law.

Composition analysis indicated that the Fe composition ratio increased after annealing. The XRD analysis showed that the amount of LaAs impurity phase also increased after annealing. These experimental results can be qualitatively explained by assuming a chemical reaction formula of LaFe$_{1.8}$As$_2$ → 0.9LaFe$_2$As$_2$ + 0.1LaAs + 0.1As. $v_{Fe}$ in the La122 phase decreases during HP synthesis because a larger amount of As forms dimers due to a compression in As-As bond length ($l_{As-As}$). The Fe deficiency in CT-La122 must be introduced to avoid an abnormal decrease in $v_{Fe}$. The larger residual resistivity of CT-La122 is explained by the scattering of carriers due to the Fe deficiency in the Fe$_2$As$_2$ conduction layers. The large $RRR$ of UT-La122, which is not inferior to other typical polycrystalline samples of iron arsenides,[1-9] shows that there is no serious Fe deficiency in the Fe$_2$As$_2$ conduction layers.

We attempted Rietveld refinements with the occupancy of the Fe site as a variable parameter. However, meaningful differences in Fe occupancy between CT-La122 and UT-La122 were not obtained. This is because the difference in the Fe occupancy is smaller than the accuracy (~10%) that can be determined by a powder diffraction experiment and Rietveld analysis. However, considering results from the composition and XRD analyses, as well as the resistivity measurements, it is reasonable to conclude that a Fe deficiency exists in CT-La122.

Comparison of the superconductivity, crystal structure, and Fermi surface topology with related compounds is interesting to explore the relationship between them. $T_c$ of UT-La122 (~12 K) is substantially lower than it is for hole-doped 122-type iron arsenides such as ($Ae$,$A$)122[9,41] and ($Ln$,Na)122 (= 25–38 K)[25,28]. However, values of $\alpha_{As-Fe-As}$ (= 110.8(2)°) and $h_{As}$ (= 1.357(2)) are close to the optimal values (109° and 1.38 Å, respectively[38,39]) obtained empirically for iron-based superconductors. One possible reason for the large difference in $T_c$ is Fermi surface topology. Cylindrical Fermi surfaces at $\Gamma$ and $X$ points play important roles in producing high-$T_c$ in iron-based superconductors. Namely, an $s\pm$ wave pairing is realized by the spin-fluctuation modes arising from the nesting across these the Fermi surfaces.[42,43] However, a hole-like Fermi surface cylinder is missing in UT-La122 due to the heavy electron-doping, which may result in the low $T_c$ of UT-La122.

It is also interesting to compare the Fermi surface topology of La122 with those of the 122-type iron selenides $A_x$Fe$_{2-y}$Se$_2$ in which the hole-like Fermi surface cylinder disappears when electrons are heavily doped.[44–46]

The analogous compound of LaFe$_2$P$_2$ does not show superconductivity down to 1.8 K,[47] although its Fermi surface topology is similar to that of UT-La122. Judging from the P-P bond length (= 3.176 Å), LaFe$_2$P$_2$ is an uncollapsed tetragonal structure and has the same $v_{Fe}$ as UT-La122. The lack of superconductivity (or low $T_c$ < 1.8 K) in LaFe$_2$P$_2$ may be due to large $\alpha_{As-Fe-As}$ (117.8°) and short $h_{As}$ (1.16 Å), which are disadvantageous for superconductivity.

$T_c$ of La122 is nearly the same as $T_c$ of Ca122 ($T_c$ ~ 12 K), which appears in either the vicinity of the collapsed tetragonal phase transition or in the tetragonal phase under pressure.[26,48-50] The Fermi surface topology of each is also similar. $v_{Fe}$ in Ca122 changes from +2 to +1 due to the structure collapse (dimerization of As). Looking at the behavior of the $T$-dependent electrical resistivity in Ca122,[49] it appears that $v_{Fe}$ gradually decreases with increasing pressure and superconductivity occurs for a certain $v_{Fe}$ between +2 and +1. Taking the similarities between Ca122 and La122 into account, superconductivity in Ca122 may occur when the $v_{Fe}$ in the tetragonal Ca122 is similar to that in La122 (~ +1.5), due to the partial dimerization of As under pressure. In other words, La122 can realize superconductivity at normal pressure, similar to Ca122 under pressure.

According to the above scenario, there should be only one superconducting phase near the La122 composition in the electron-doped Ca122 and possibly (La,Na)122. If so, their electronic phase diagrams are different from those of (Sr,La)122 and (Ba,La)122.[17,18] Although we have not succeeded in synthesizing electron-doped samples except for LaFe$_2$As$_2$, an interesting avenue of future work would be to investigate the other electron-doping region of (La,Na)122 and/or La-doped Ca122 in order to establish its entire electronic phase diagram and compare them with other systems.

In summary, we succeeded in synthesizing La122 using an HP technique for the first time. We found that the structure of La122 changes from a collapsed to uncollapsed tetragonal through annealing, and superconductivity emerged at 12.1 K in the uncollapsed tetragonal phase. La122 can provide a platform to study heavily electron-doped iron-based superconductors. Additional physical property measurements and theoretical studies on La122 are expected to introduce curious aspects to the physics and chemistry of iron-based superconductors.

## AUTHOR INFORMATION
### Corresponding Author
Akira Iyo
E-mail: iyo-akira@aist.go.jp